\documentclass[journal,12pt,onecolumn]{IEEEtran}

\usepackage{amsmath,amssymb,epsfig,psfrag,cite,subfigure}
\include{macros}
\usepackage{graphicx,widetext}

\usepackage{algorithm}
\usepackage{epsfig,psfrag}
\usepackage{subfigure}
\usepackage{color}
\usepackage{url}
\DeclareMathOperator*{\argmin}{arg\,min}
\DeclareMathOperator*{\argmax}{arg\,max}

\usepackage{epstopdf}
\newcommand*{\QEDA}{\hfill\ensuremath{\blacksquare}}%

\begin{document}

\title{Hybrid TDOA/RSS Based Localization for\\Visible Light Systems}

\author{Ertan Kazikli\thanks{The authors are with the Department of Electrical and Electronics Engineering, Bilkent University, 06800, Ankara, Turkey, Tel: +90-312-290-3139, Fax: +90-312-266-4192, Emails: \{kazikli,gezici\}@ee.bilkent.edu.tr.} and Sinan Gezici\vspace{-1.2cm}}

% make the title area
\maketitle
%\thispagestyle{empty}
%\pagestyle{empty}

%Impact Statement:
%For the first time in the literature, the joint utilization of received signal strength (RSS) and time difference of arrival (TDOA) parameters is proposed for visible light positioning (VLP) systems. Theoretical limits on localization accuracy are quantified by deriving the Cramer-Rao lower bound (CRLB) expression for a generic quasi-synchronous VLP system. Both direct positioning and two-step positioning approaches are developed for location estimation in quasi-synchronous VLP systems. It is shown that the proposed estimators can achieve the CRLB at high signal-to-noise ratios (SNRs).

\begin{abstract}
In a visible light positioning (VLP) system, a receiver can estimate its location based on signals transmitted by light emitting diodes (LEDs). In this manuscript, we investigate a quasi-synchronous VLP system, in which the LED transmitters are synchronous among themselves but are not synchronized with the receiver. In quasi-synchronous VLP systems, position estimation can be performed by utilizing time difference of arrival (TDOA) information together with channel attenuation information, leading to a hybrid localization system. To specify accuracy limits for quasi-synchronous VLP systems, the Cram\'{e}r-Rao lower bound (CRLB) on position estimation is derived in a generic three-dimensional scenario. Then, a direct positioning approach is adopted to obtain the maximum likelihood (ML) position estimator based directly on received signals from LED transmitters. In addition, a two-step position estimator is proposed, where TDOA and received signal strength (RSS) estimates are obtained in the first step and the position estimation is performed, based on the TDOA and RSS estimates, in the second step. The performance of the two-step positioning technique is shown to converge to that of direct positioning at high signal-to-noise ratios based on asymptotic properties of ML estimation. Finally, CRLBs and performance of the proposed positioning techniques are investigated through simulations.

\textit{Keywords:} Estimation, received signal strength (RSS), time difference of arrival (TDOA), localization, visible light.
\end{abstract}

\section{Introduction}\label{sec:intro}

Recently, there has been a growing interest in the potential use of visible light systems based on light emitting diodes (LEDs) for the purpose of communications, especially in indoor environments \cite{HighSpeedVLC,VLCBeyond,LEDBasedIndoor,VLCNetworkingSensing}.
Since LEDs are increasingly deployed for illumination purposes due to their energy efficiency, integration of visible light communication (VLC) to LED networks appears as an appealing idea to provide communication and illumination simultaneously \cite{HighSpeedVLC,VLCBeyond,LEDBasedIndoor,VLCNetworkingSensing,IlluminationPersVLC}.
The potential widespread use of LEDs also inspires a growing number of visible light positioning (VLP) systems, in which signals transmitted by LEDs are utilized for location estimation
\cite{IndoorPosVLCSurvey,LetLightGuideUs,epsilon,VLP_CRLB_RSS,PosAccuracyTOA,VLP_CRLB_RSS_2017}. LED based localization is a promising approach as it can provide highly accurate position information inexpensively through installation of a few LEDs \cite{Keskin_Direct,VLP_tilted,zhang2014asynchronous,TDOABasedLED}, which is beneficial for various applications such as asset tracking and robotic control \cite{VLP_Roadmap}.

Similar to radio frequency (RF) based localization, VLP systems utilize various parameters such as time of arrival (TOA), time difference of arrival (TDOA), received signal strength (RSS), and/or angle of arrival (AOA) for extracting the position of a target object (i.e., a VLC receiver) \cite{IndoorPosVLCSurvey,Sinan_Survey}. In practice, the choice of parameters to be employed for localization is determined based on desired accuracy levels and system requirements. An important system requirement is the presence of a synchronization mechanism, which is necessary for time based VLP systems that utilize TOA or TDOA information. Depending on the existence of a synchronization mechanism, VLP systems can be categorized as \textit{synchronous}, \textit{asynchronous}, and \textit{quasi-synchronous}. In a synchronous VLP system, all LED transmitters and VLC receivers are synchronized (for example, via a common reference clock) whereas there is no synchronization among any of them in an asynchronous VLP system. On the other hand, in quasi-synchronous VLP systems, LED transmitters are synchronized among themselves but are asynchronous with VLC receivers. Asynchronous VLP systems facilitate low-complexity implementations whereas the synchronous VLP systems have the highest complexity. Between these two categories, quasi-synchronous systems require synchronization only among LED transmitters, which can be realized relatively easily via cabling during the installation of LED infrastructures.

In synchronous VLP systems, location related information can be extracted from the TOA parameter based on its relation to the time-of-flight of a received signal \cite{PosAccuracyTOA,comparative_keskin,ZZB_MFK,Keskin_Direct}. In \cite{PosAccuracyTOA}, the time delay parameter of a received signal is exploited to perform range (i.e., distance) estimation in a synchronous VLP system, and the corresponding Cram\'{e}r-Rao lower bound (CRLB) is investigated for various system parameters. In addition, the Ziv-Zakai bound (ZZB) for range estimation is derived for synchronous VLP systems in the presence of prior information in \cite{ZZB_MFK}. Moreover, a synchronous VLP system employing both TOA and RSS information is investigated in \cite{Keskin_Direct}. This investigation includes not only a theoretical framework, which provides a CRLB expression for position estimation in a generic three-dimensional scenario, but also direct and two-step estimation algorithms for extracting the position of a VLC receiver, which are shown to achieve accuracies as high as the theoretical limit for high signal-to-noise ratios (SNRs).

Due to its low-complexity nature, asynchronous VLP systems have been considered in numerous papers in the literature such as
\cite{epsilon,zhang2014asynchronous,VLP_CRLB_RSS,AOA_Kalman,AOA_RSS,BiasedDistEst,Hybrid3D,Keskin_Direct,VLP_CRLB_RSS_2016,VLP_CRLB_RSS_2017,3DPosVLC_PSO}. For instance, the work in \cite{epsilon} employs RSS measurements to obtain the desired position via trilateration. On the other hand, a theoretical analysis is carried out to explore the fundamental limits on the achievable accuracy of RSS based position estimators in \cite{VLP_CRLB_RSS_2017}, which considers a system with multiple photodiodes placed on the target object forming an aperture-based VLC receiver. Moreover, \cite{Hybrid3D} combines AOA and RSS information to enhance positioning accuracy in an asynchronous VLP system and illustrates performance improvements over AOA based positioning via simulations. Finally, \cite{Keskin_Direct} investigates an asynchronous VLP systems by providing theoretical results on attainable accuracies as well as algorithms for estimating the desired position.

% In quasi-synchronous VLP systems, the relative travel time information (i.e., TDOA) of transmitted signals from a set of LEDs can be utilized by a VLC receiver since LEDs are synchronous among themselves. The work in \cite{TDOABasedLED} focuses on an LED based localization system in which a VLC receiver is located, with centimeter level accuracy, based on TDOA measurements. Although \cite{TDOABasedLED} presents a practical localization algorithm for quasi-synchronous VLP systems, theoretical limits and optimal estimators have not been investigated for such systems in the literature. In addition, joint utilization of TDOA and RSS information has not been considered for quasi-synchronous VLP systems. Even though some papers, such as \cite{TDOA_RSS_Wireless,TDOA_RSS_Wireless2}, focus on hybrid positioning schemes that employ both TDOA and RSS parameters in RF based positioning systems, localization in visible light systems requires new formulations and analyses as the channel characteristics are significantly different in optical systems compared to those in RF systems.

{In quasi-synchronous VLP systems, the relative travel time information (i.e., TDOA) of transmitted signals from a set of LEDs can be utilized by a VLC receiver since LEDs are synchronous among themselves. Various studies in the literature utilize the TDOA parameter for position estimation \cite{TDOABasedLED,TDOAPhotonicsJournal2018,TDOANaeem2018,TDOADo2014,TDOAChen2018,TDOACoherentHeterodyne2013,TDOASinusoidal2013,TDOASingleLED2018,BirendraOFDM}. For instance, \cite{TDOANaeem2018} investigates theoretical accuracy limits for position estimation based on TDOA measurements. The work in \cite{TDOABasedLED} focuses on an LED based localization system in which a VLC receiver is located with centimeter level accuracy by employing TDOA measurements. A recent study in \cite{TDOAPhotonicsJournal2018} proposes a practical low complexity VLP system implemented on a hardware which utilizes TDOA parameters and reports the positioning accuracy as $9.2$ centimeters. Although there exist studies focusing on practical localization algorithms for quasi-synchronous VLP systems, theoretical limits and optimal estimators have not been investigated for such systems in the literature. In addition, joint utilization of TDOA and RSS information has not been considered for quasi-synchronous VLP systems. Even though some papers, such as \cite{TDOA_RSS_Wireless,TDOA_RSS_Wireless2}, focus on hybrid positioning schemes that employ both TDOA and RSS parameters in RF based positioning systems, localization in visible light systems requires new formulations and analyses as the channel characteristics are significantly different in optical systems compared to those in RF systems.}

% Although there exist many studies that investigate either synchronous or asynchronous VLP systems, quasi-synchronous VLP systems are rarely discussed in the literature \cite{TDOABasedLED}.
Since VLC receivers are commonly placed on mobile objects, cabling is not an option for synchronizing LED transmitters with VLC receivers; hence, realizing a synchronous VLP system (with precise synchronization) is costly and challenging. On the other hand, the quasi-synchronous scenario requires a synchronization mechanism only among LED transmitters, which are usually at fixed locations (e.g., on the ceiling of a room). Therefore, quasi-synchronous VLP systems are practical and cost effective compared to synchronous VLP systems. In addition, they enable the use of the TDOA parameter, which cannot be utilized in asynchronous VLP systems, to extract location related information. Overall, it is important to investigate quasi-synchronous VLP systems, which is the aim of this manuscript.

In this work, quasi-synchronous VLP systems, which utilize both TDOA and RSS information, are analyzed. In particular, a CRLB expression is derived for position estimation in such systems. To the best of authors' knowledge, theoretical limits for quasi-synchronous VLP systems have not been available in the literature. The provided CRLB expression is generic in the sense that it is valid for any system parameters such as orientations of LED transmitters and shapes of pulses transmitted from LEDs. Moreover, the maximum likelihood (ML) estimator for the position of a VLC receiver is obtained considering a direct positioning approach, in which position estimation is performed based on received signals directly. Furthermore, a two-step estimator is proposed relying on the asymptotic properties of ML estimation. For the first time in the literature, a two-step positioning technique for a quasi-synchronous VLP system, which utilizes both TDOA and RSS parameters, is developed. It is shown that the performance of two-step positioning, which is computationally less demanding than direct positioning, approaches that of the direct positioning approach at high SNRs.

The remainder of this manuscript is organized as follows. Section~\ref{sec:sysmodel} describes the considered VLP system. Section~\ref{sec:crlb} consists of the derivation of the CRLB for localization in quasi-synchronous VLP systems. Section~\ref{sec:mle} investigates ML based positioning techniques, namely, direct positioning and two-step positioning. Section~\ref{sec:nume} presents numerical examples to illustrate both the theoretical limits and the performance of the proposed estimators. Finally, concluding remarks are made in Section~\ref{sec:conc}.

%=============================================================================
\section{System Model}\label{sec:sysmodel}

In the considered VLP system,
LED based transmitters and a photo detector based VLC receiver are employed to localize a target object in an indoor environment.
In particular, the LED transmitters are placed at various locations in a room (e.g., on the ceiling)  and the photo detector based VLC receiver is placed on the target object. Each of the LED transmitters emits a known visible light signal. It is assumed that only LOS components of the transmitted signals are received by the VLC receiver at the target object, which aims to estimate its own position.

The locations of the LED transmitters are known and denoted by $\boldsymbol{l}_t^i\in\mathbb{R}^3$ for $i=1,\dots,N_L$, where $N_L$ stands for the number of LED transmitters. The aim is to estimate the unknown location of the VLC receiver, denoted by $\boldsymbol{l}_r\in\mathbb{R}^3$, based on the signals coming from the LED transmitters. The received signal at the VLC receiver due to the transmission from the $i$th LED transmitter is modeled by \cite{PosAccuracyTOA,comparative_keskin}
\begin{align}\label{eq:sigmodel}
r_{i}(t) = \alpha_i R_p s_i(t-\tau_i) + n_i(t)
\end{align}
for $t\in [T_1^i,T_2^i]$, where $T_1^i$ and $T_2^i$ denote, respectively, the initial and final instants of the observation interval for the reception of the signal coming from the $i$th LED transmitter, $\alpha_i$ is the attenuation factor of the optical channel between the $i$th LED transmitter and the VLC receiver ($\alpha_i>0$), $R_p$ is the responsivity of the photo detector, $s_i(t)$ is the transmitted signal from the $i$th LED transmitter (which is nonzero over an interval of $[0,T_s]$), $\tau_i$ is the time of arrival (TOA) parameter of the signal transmitted from the $i$th LED transmitter, and $n_i(t)$ is zero-mean additive white Gaussian noise (AWGN) with a power spectral density level of $\sigma^2$. It is assumed in this model that the signals coming from different LED transmitters do not interfere with each other, which in practice can be achieved by using such multiple access techniques as time division multiplexing or frequency division multiplexing \cite{multiaccessVLP}.\footnote{In the case of time division multiplexing, it is also assumed that the position of the VLC receiver stays the same over different time slots.}
In addition, $n_i(t)$ and $n_j(t)$ are modeled to be independent for $i\neq j$ as they are observed over different time or frequency intervals (due to time or frequency division multiplexing).
% Moreover, it is assumed that the VLC receiver knows $R_p$ and $s_i(t)$ and can use this information during the localization process \cite{Keskin_Direct,Hybrid3D}.

In the considered setting, the LED transmitters are synchronous with each other while they are asynchronous with the VLC receiver, which corresponds to a quasi-synchronous scenario. As the LED transmitters are placed at fixed locations in the room, it is easy to synchronize their clocks for instance via a wired synchronization system. However, the VLC receiver is not necessarily at a fixed location; hence, it is difficult to synchronize its clock with those of the LED transmitters. Therefore, the considered scenario is commonly encountered in practical applications. Under this setting, the TOA parameter of the signal coming from the $i$th LED transmitter can be expressed as
\begin{align}\label{eq:toadef}
\tau_i = \frac{\lVert \boldsymbol{l}_r - \boldsymbol{l}_t^i \rVert}{c} + \Delta\,,
\end{align}
where $\lVert \cdot \rVert$ denotes the Euclidean norm, $c$ is the speed of light, and $\Delta$ is the time offset between the clocks of an LED transmitter and the VLC receiver. Note that this time offset is the same for all LED transmitters as they are synchronous with each other. Moreover, $\Delta$ is modeled as a deterministic unknown parameter since it takes a fixed value which is unknown to the localization process. Furthermore, it is assumed that the signal component in \eqref{eq:sigmodel} is fully captured at the VLC receiver by having an appropriate observation interval for the reception.

For synchronized LED transmitters that are asynchronous with the VLC receiver, the TDOA parameter can be utilized for localization \cite{Sinan_Survey}.
One way of generating TDOA measurements is to select one of the LED transmitters as the reference and to compute the TDOA parameters of the signals coming from the other LED transmitters with respect to the reference. Thus, the TDOA parameter of the signal coming from the $i$th LED transmitter can be expressed as
\begin{align}\label{eq:tdoadef}
d_i = \tau_i - \tau_{1}\,,
\end{align}
for $i\in\{2,\dots,N_L\}$, where the first LED transmitter is chosen as the reference for notational convenience. It is worth noting that as the same $\Delta$ is present as an additive term in all the $\tau_i$ parameters, the resulting $d_i$ does not contain the time offset parameter.

The attenuation factor of the optical channel between the $i$th LED transmitter and the VLC receiver is modeled as
\begin{align}\label{eq:alphadef}
\alpha_i =
\frac{m_i+1}{2\pi}
\cos^{m_i}(\phi_i)
\cos(\theta_i)
\frac{A_r}{\lVert \boldsymbol{l}_r-\boldsymbol{l}_t^i\rVert^2}\,,
\end{align}
where $m_i$ is the Lambertian order of the $i$th LED transmitter, $\phi_i$ and $\theta_i$ are, respectively, the irradiation and the incidence angles for the channel between the $i$th LED transmitter and the VLC receiver, and $A_r$ is the area of the photo detector \cite{PosAccuracyTOA,epsilon,comparative_keskin}. Note that $\alpha_i$ is also referred to as the received signal strength (RSS) parameter as it directly determines the received signal power at the VLC receiver. We can also define normal vectors $\boldsymbol{n}_t^i \in \mathbb{R}^3$ and $\boldsymbol{n}_r\in \mathbb{R}^3$ as the directions of the $i$th LED transmitter and the VLC receiver, respectively, to express the attenuation factor of the $i$th channel in the following form:
\begin{align}\label{eq:alphadef2}
\alpha_i =
\gamma_i
\frac{
\left(
(\boldsymbol{l}_r-\boldsymbol{l}_t^i)^T\boldsymbol{n}_t^i
\right)^{m_i}
\left(
(\boldsymbol{l}_t^i-\boldsymbol{l}_r)^T\boldsymbol{n}_r
\right)
}{
\lVert \boldsymbol{l}_r - \boldsymbol{l}_t^i \rVert^{m_i+3}
}\,,
\end{align}
where $\gamma_i\triangleq (m_i+1)A_r/(2\pi)$. The equivalent expression in \eqref{eq:alphadef2} is helpful in the following derivations as the full dependency of $\alpha_i$ on $\boldsymbol{l}_r$ is shown explicitly. {In the specified system model, it is assumed that the VLC receiver knows $R_p$, $A_r$, $\boldsymbol{n}_r$, $s_i(t)$, $m_i$, $\boldsymbol{n}_t^i$ and $\boldsymbol{l}_t^i$ for $i=1,\dots,N_L$ and can use this information during the localization process \cite{Keskin_Direct,Hybrid3D}. In other words, the only unknown parameters are the position of the VLC receiver $\boldsymbol{l}_r$ and the time offset $\Delta$.}

%=============================================================================
\section{Theoretical Limits}\label{sec:crlb}

In this section, theoretical limits on localization accuracy are investigated for the quasi-synchronous VLP system model described in the previous section. In particular, the derivation of the CRLB is presented for estimating the unknown parameters, which consist of the position of the VLC receiver as well as the time offset between the clocks of an LED transmitter and the VLC receiver.

Considering the received signal model in \eqref{eq:sigmodel} and observing that $n_i(t)$ and $n_j(t)$ are independent for $i\neq j$, the log-likelihood function is given by
\begin{align}\label{eq:loglike}
\Lambda(\boldsymbol{\varphi}) =
k - \frac{1}{2\sigma^2}
\sum_{i=1}^{N_L}
\int_{T_1^i}^{T_2^i} ( r_i(t) - \alpha_i R_p s_i(t-\tau_i) )^2 dt
\end{align}
where $\boldsymbol{\varphi}=[ \boldsymbol{l}_r^T,\Delta]^T\in\mathbb{R}^4$ represents the unknown parameter vector and $k$ is a normalizing constant which does not depend on $\boldsymbol{\varphi}$ \cite{geolocation,ziv_zakai_delay}. The computation of the CRLB is performed as follows: First, the Fisher information matrix (FIM) is obtained based on the log-likelihood function in \eqref{eq:loglike} as \cite{poor_book}
\begin{align}\label{eq:fim}
\mathbf{J} (\boldsymbol{\varphi} ) =
\mathbb{E}
\left\{
(\nabla_{\boldsymbol{\varphi}}  \Lambda(\boldsymbol{\varphi} ))
(\nabla_{\boldsymbol{\varphi}}  \Lambda(\boldsymbol{\varphi} ))^T
\right\}
\end{align}
where $\nabla_{\boldsymbol{\varphi} }\Lambda(\boldsymbol{\varphi} )$ is the gradient vector of the log-likelihood function with respect to the unknown parameter vector. The next step is to take the inverse of the FIM in order to express the CRLB on the covariance matrix of any unbiased estimator $\boldsymbol{\hat{\varphi}}$ of $\boldsymbol{\varphi}$ as
\begin{align}\label{eq:crlb}
\mathbb{E}\{ (\boldsymbol{\hat{\varphi} } - \boldsymbol{\varphi} ) (\boldsymbol{\hat{\varphi} } - \boldsymbol{\varphi} )^T \}
\succeq \mathbf{J}(\boldsymbol{\varphi} )^{-1}
\end{align}
where $\mathbf{A}\succeq \mathbf{B}$ means that $\mathbf{A}-\mathbf{B}$ is positive semidefinite \cite{poor_book}. By focusing only on the diagonal terms in \eqref{eq:crlb}, one can also write
\begin{align}\label{eq:crlb2}
\mathrm{Var}(\hat{\varphi}_k) \geq [\mathbf{J}(\boldsymbol{\varphi})^{-1}]_{k,k}
\end{align}
where $\hat{\varphi}_k$ is the $k$th entry of $\boldsymbol{\hat{\varphi}}$ and $[\,\cdot\,]_{k,k}$ denotes the $k$th diagonal entry of its argument.
{It is noted that the FIM matrix for the synchronous VLP system (i.e., known $\Delta$) is derived in \cite[Prop.~1]{Keskin_Direct}. The FIM matrix for the quasi-synchronous VLP system considered in this study can be found by extending the FIM matrix derived in \cite{Keskin_Direct}. In particular, the elements of the FIM in \eqref{eq:fim} can be obtained from the log-likelihood function in \eqref{eq:loglike} after some manipulation (please see Appendix~\ref{sec:fimder} for details) as
\begin{align}\label{eq:Jmn}
[\mathbf{J}(\boldsymbol{\varphi})]_{m,n}
=\frac{R_p^2}{\sigma^2}
\sum_{i=1}^{N_L}
\bigg(E_2^i
\frac{\partial \alpha_i}{\partial l_{r,m}}
\frac{\partial \alpha_i}{\partial l_{r,n}}
+\alpha_i^2 E_1^i
\frac{\partial \tau_i}{\partial l_{r,m}}
\frac{\partial \tau_i}{\partial l_{r,n}}
-\alpha_i E_3^i\Big(\frac{\partial \alpha_i}{\partial l_{r,m}}
\frac{\partial \tau_i}{\partial l_{r,n}}
+\frac{\partial \tau_i}{\partial l_{r,m}}
\frac{\partial \alpha_i}{\partial l_{r,n}}
\Big)
\bigg)
\end{align}
for $m,n=1,2,3$,
\begin{align}\label{eq:Jk4}
[\mathbf{J}(\boldsymbol{\varphi})]_{4,k}
=[\mathbf{J}(\boldsymbol{\varphi})]_{k,4}
=\frac{R_p^2}{\sigma^2}
\sum_{i=1}^{N_L}
\bigg(
\alpha_i^2 E_1^i \frac{\partial\tau_i}{\partial l_{r,k}}
-\alpha_i E_3^i \frac{\partial\alpha_i}{\partial l_{r,k}}
\bigg)
\end{align}
for $k=1,2,3$, and
\begin{align}\label{eq:J44}
[\mathbf{J}(\boldsymbol{\varphi})]_{4,4}
= \frac{R_p^2}{\sigma^2}
\sum_{i=1}^{N_L} \alpha_i^2 E_1^i\,,
\end{align}
where $l_{r,k}$ denotes the $k$th element of $\boldsymbol{l}_r$, the integrals involving $s_i(t)$ and the derivative of $s_i(t)$, denoted by $s_i'(t)$, are defined as
\begin{align}\label{eq:e1}
E_1^i
&\triangleq \int_{0}^{T_s}
s_i'(t)^2 dt,\\\label{eq:e2}
E_2^i
&\triangleq \int_{0}^{T_s}
s_i(t)^2 dt,\\\label{eq:e3}
E_3^i
&\triangleq \int_{0}^{T_s}
s_i(t)s'_i(t)dt
\end{align}
and the partial derivatives in \eqref{eq:Jmn} and \eqref{eq:Jk4} of $\alpha_i$ and $\tau_i$ with respect to the coordinates of the VLC receiver position are as in \cite[Prop.~1]{Keskin_Direct}.}
%\begin{align}\label{eq:derTau}
%\frac{\partial\tau_i}{\partial l_{r,k}} =
%\frac{(l_{r,k}-l_{t,k}^i)}
%{c
%\lVert \boldsymbol{l}_r - \boldsymbol{l}_t^i \rVert
%}
%\end{align}
%and
%\begin{align}
%\frac{\partial\alpha_i}{\partial l_{r,k}}
%=
%\frac{
%\gamma_i m_i n_{t,k}^i
%\left(
%(\boldsymbol{l}_r-\boldsymbol{l}_t^i)^T\boldsymbol{n}_t^i
%\right)^{m_i-1}
%\left(
%(\boldsymbol{l}_t^i-\boldsymbol{l}_r)^T\boldsymbol{n}_r
%\right)
%}{
%\lVert \boldsymbol{l}_r - \boldsymbol{l}_t^i \rVert^{m_i+3}
%}
%-\frac{
%\gamma_i
%n_{r,k}
%\left(
%(\boldsymbol{l}_r-\boldsymbol{l}_t^i)^T\boldsymbol{n}_t^i
%\right)^{m_i}
%}{
%\lVert \boldsymbol{l}_r - \boldsymbol{l}_t^i \rVert^{m_i+3}
%}+
%\frac{\alpha_i (l_{t,k}^i-l_{r,k})(m_i+3)}
%{\lVert \boldsymbol{l}_r - \boldsymbol{l}_t^i \rVert^2}
%\end{align}
%with $n_{r,k}$, $n_{t,k}^i$ and $l_{t,k}^i$ denoting the $k$th entries of $\boldsymbol{n}_r$, $\boldsymbol{n}_t^i$, and $\boldsymbol{l}_t^i$,
%respectively.

{
\textit{\textbf{Remark~1}:} It should be emphasized that the derived CRLB expressions provide bounds on variances of unbiased estimators. For biased estimators, theoretical limits on positioning accuracy are in general different from the ones derived in this manuscript. However, if the form of the bias is known, the results in this study can be extended to provide a bound on the achievable accuracy of such biased estimators by using the information inequality \cite[p.~169]{poor_book}.
}

{It should be noted that the FIM matrix for the considered quasi-synchronous VLP system is $4\times 4$ whereas the FIM matrix for the synchronous VLP system \cite{Keskin_Direct} is $3\times 3$. Moreover, all the entries of the FIM matrix of the synchronous VLP system appear in the FIM matrix of the quasi-synchronous VLP system (i.e., $J_{m,n}$ in \eqref{eq:Jmn} for $m,n\in\{1,2,3\}$ correspond to the entries of the synchronous VLP system). On the other hand, the entries specified in \eqref{eq:Jk4} and \eqref{eq:J44} are the additional terms for the quasi-synchronous VLP system which is due to the fact that $\Delta$ is an unknown parameter in this case.}

After obtaining the FIM, one can simply take its inverse to compute the CRLB for estimating the position of the VLC receiver (see \eqref{eq:crlb} and \eqref{eq:crlb2}). As a result, the lower bound on the localization accuracy can be assessed by computing the CRLB for any given system configuration. In fact, based on the following proposition, the computational complexity of the CRLB calculation can be reduced.

\textit{\textbf{Proposition~1}:}
The CRLB on the MSE of any unbiased estimator $\hat{\boldsymbol{l}}_r$ of $\boldsymbol{l}_r$ can be expressed as
\begin{align}\label{eq:crlbmse}
\mathbb{E}\{
\lVert \boldsymbol{l}_r - \hat{\boldsymbol{l}}_r \rVert^2
\}
\geq
\mathrm{trace}\left\{
\mathbf{J}_{\rm{qs}}^{-1}
\right\}
\end{align}
where $\mathbf{J}_{\rm{qs}}$ represents a $3\times 3$ matrix with the following entries:
\begin{align}
[\mathbf{J}_{\rm{qs}}]_{m,n}&=
\frac{R_p^2}{\sigma^2\sum_{i=1}^{N_L} \alpha_i^2 E_1^i}
\sum_{i=1}^{N_L}
\sum_{j=1}^{N_L}
\frac{\partial \alpha_i}{\partial l_{r,m}}
\bigg(\alpha_j^2 E_2^i E_1^j
\frac{\partial \alpha_i}{\partial l_{r,n}}
-\alpha_i \alpha_j E_3^i E_3^j
\frac{\partial\alpha_j}{\partial l_{r,n}}
+\alpha_i \alpha_j^2 E_3^i E_1^j
\Big(\frac{\partial\tau_j}{\partial l_{r,n}}-
\frac{\partial \tau_i}{\partial l_{r,n}}
\Big)
\bigg)
\nonumber\\\label{eq:blockinv1}
&+\frac{\partial \tau_i}{\partial l_{r,m}}
\bigg(
\alpha_j^2 E_1^j
\Big(
\alpha_i^2 E_1^i
\frac{\partial \tau_i}{\partial l_{r,n}}
-\alpha_i E_3^i
\frac{\partial \alpha_i}{\partial l_{r,n}}
\Big)
+\alpha_i^2 E_1^i
\Big(\alpha_j E_3^j\frac{\partial\alpha_j}{\partial l_{r,n}}
-\alpha_j^2 E_1^j
\frac{\partial\tau_j}{\partial l_{r,n}}
\Big)\bigg)
\end{align}
for $m,n\in \{1,2,3\}$.

\textit{Proof:} Let the FIM in \eqref{eq:crlb} be partitioned as
\begin{align}\label{eq:fimpartition}
\mathbf{J}(\boldsymbol{\varphi}) =
\begin{bmatrix}
\mathbf{J}_{\mathbf{A}} & \mathbf{J}_\mathbf{b}\\
\mathbf{J}_\mathbf{b}^T & {\rm{J}_c}
\end{bmatrix}
\end{align}
where $\mathbf{J}_{\mathbf{A}}$ is a $3\times3$ matrix specified by \eqref{eq:Jmn},  $\mathbf{J}_{\mathbf{b}}$ is a $3\times1$ vector with entries specified by \eqref{eq:Jk4}, and ${\rm{J}_c}$ is a scalar given by \eqref{eq:J44}. Then, the entries of the inverse FIM that are related to the estimation of the VLC receiver position only can be expressed as
\begin{align}\label{eq:blockinv2}
\left[\mathbf{J}(\boldsymbol{\varphi})^{-1}\right]_{3\times 3} =
\left(
\mathbf{J}_{\mathbf{A}} - \frac{1}{{\rm{J}_c}}\, \mathbf{J}_\mathbf{b}\mathbf{J}_\mathbf{b}^T
\right)^{-1}.
\end{align}
By plugging the expressions in \eqref{eq:Jmn}, \eqref{eq:Jk4}, and \eqref{eq:J44} into \eqref{eq:blockinv2} and after some manipulation, the expressions in \eqref{eq:crlbmse} and \eqref{eq:blockinv1} can be obtained via \eqref{eq:crlb}.
\QEDA

The result in Proposition~1 is important as it gives an alternative and equivalent way of calculating the CRLB on the MSE of any unbiased estimator for the position of the VLC receiver. It is noted that the expression in \eqref{eq:crlbmse} requires a $3\times 3$ matrix inversion while the original expression in \eqref{eq:crlb} leads to a $4\times 4$ matrix inversion.

% It is important to note that $\mathbf{J}_{\mathbf{A}}$ defined in \eqref{eq:fimpartition} corresponds to the FIM in the case of a synchronous VLP system, equivalently, in the case of known $\Delta$ \cite{Keskin_Direct}.
{As mentioned earlier, $\mathbf{J}_{\mathbf{A}}$ defined in \eqref{eq:fimpartition} corresponds to the FIM in the case of a synchronous VLP system, equivalently, in the case of known $\Delta$ \cite{Keskin_Direct}. Therefore, the additional unknown parameter $\Delta$ in the case of a quasi-synchronous VLP system leads to the second term on the right hand side of \eqref{eq:blockinv2} that is subtracted from the FIM of the synchronous VLP system, i.e., $\mathbf{J}_{\mathbf{A}}$, while obtaining the bound on the MSE of unbiased position estimators for quasi-synchronous VLP systems.}
In addition, when the elements of $\mathbf{J}_{\mathbf{b}}$ in \eqref{eq:fimpartition} are zero, the CRLB for the synchronous VLP system becomes identical to that for the quasi-synchronous VLP system, that is, $[\mathbf{J}(\boldsymbol{\varphi})^{-1}]_{3\times 3}= \mathbf{J}_{\mathbf{A}}^{-1}$, as can also be observed from \eqref{eq:blockinv2}. In particular, when $E_3^i=0$ for $i=1,\dots,N_L$ (which is the case for common pulses in practice), synchronous and quasi-synchronous VLP systems have the same theoretical limits if the following conditions hold (see \eqref{eq:Jk4}):
\begin{align}\label{eq:blockdiagcond}
\sum_{i=1}^{N_L} \alpha_i^2 E_1^i \frac{(l_{r,k}-l_{t,k}^i)}{\lVert \boldsymbol{l}_r-\boldsymbol{l}_{t,k}^i\rVert} = 0
\end{align}
for all $k=1,2,3$. In this case, the CRLB does not depend on whether $\Delta$ is known or unknown. However, the conditions in \eqref{eq:blockdiagcond} may not hold in most cases since they require specific symmetry conditions.

%=============================================================================================
\section{Direct and Two-Step Estimators}\label{sec:mle}

In this section, ML based estimators\footnote{{It is known that the ML estimator is asymptotically unbiased and efficient \cite[p.~183]{poor_book}}.} are developed for the localization of the VLC receiver; i.e., for estimating $\boldsymbol{l}_r$. In particular, both direct positioning and two-step positioning approaches are proposed.

\subsection{Direct Positioning}

Considering the log-likelihood function in \eqref{eq:loglike}, the ML estimate of the unknown parameter vector can be expressed as
\begin{gather}\label{eq:mle}
\boldsymbol{\hat{\varphi}} =
\argmax_{\boldsymbol{\varphi}}
\sum_{i=1}^{N_L}
\alpha_i \int_{T_1^i}^{T_2^i}
r_i(t) s_i(t-\tau_i) dt
- \frac{R_p}{2} \sum_{i=1}^{N_L}
\alpha_i^2 E_2^i
\end{gather}
where $E_2^{i}$ is as defined in \eqref{eq:e2}. Then, the first three entries of $\boldsymbol{\hat{\varphi}}$ yields the ML position estimate of the VLC receiver denoted by $\boldsymbol{\hat{l}}_r$. As there exist no intermediate steps in estimating $\boldsymbol{l}_r$, this approach is referred to as direct positioning \cite{Sinan_Survey}. Note that the objective function in \eqref{eq:mle} needs to be optimized with respect to $\boldsymbol{l}_r$ and $\Delta$ jointly as they are both contained in $\boldsymbol{\varphi}$. Therefore, compared to the synchronous scenario where all the transmitters and the VLC receiver are synchronized \cite{Keskin_Direct}, $\Delta$ is an additional unknown parameter that should be estimated in this case.

\subsection{Two-Step Positioning}

The direct position estimator in \eqref{eq:mle} has high computational complexity in general as it requires a search over a four-dimensional space. For the purpose of obtaining a low-complexity estimator, a two-step position estimator is proposed in this section for localizing the VLC receiver. Two-step positioning is a common approach in the localization literature, where such position related parameters as TOA, TDOA, AOA, or RSS are estimated in the first step and then those estimated parameters are used to obtain the desired position in the second step \cite{sahinoglu:uwb,Sinan_Survey,Keskin_Direct}. In this section, a hybrid approach is proposed, which uses both TDOA and RSS measurements from the first step to obtain the position estimate of the VLC receiver in the second step. To the best of our knowledge, this is the first time that TDOA and RSS parameters are employed jointly for localization in VLP systems.

In the first step of the proposed estimator, the aim is to estimate $\tau_i$ and $\alpha_i$ related to each LED transmitter, that is, for $i=1,\dots,N_L$. Towards that aim, the log-likelihood function corresponding to the received signal due to the $i$th LED transmitter, $r_i(t)$ in \eqref{eq:sigmodel}, is maximized as follows (cf.~\eqref{eq:loglike}):
\begin{equation}\label{eq:ts_objective_init}
\{ \hat{\tau}_i,\hat{\alpha}_i\} =
\argmax_{\tau_i,\alpha_i}\,
- \frac{1}{2\sigma^2}
\int_{T_1^i}^{T_2^i} ( r_i(t) - \alpha_i R_p s_i(t-\tau_i) )^2 dt
\end{equation}
which is equivalent to (cf.~\eqref{eq:mle})
\begin{equation}\label{eq:ts_objective}
\{ \hat{\tau}_i,\hat{\alpha}_i\} =
\argmax_{\tau_i,\alpha_i}~
\alpha_i \int_{T_1^i}^{T_2^i}
r_i(t) s_i(t-\tau_i) dt
- \frac{R_p}{2} \alpha_i^2 E_2^i
\end{equation}
for $i=1,\dots,N_L$.
Similar to \cite[Section~III.C]{Keskin_Direct}, the solution of \eqref{eq:ts_objective} can be obtained as follows:
\begin{align}\label{eq:corr}
\hat{\tau}_i &= \argmax_{\tau_i}
\int_{T_1^i}^{T_2^i} r_i(t) s_i(t-\tau_i) dt\\\label{eq:alpEst}
\hat{\alpha}_i &=
\frac{C_{rs}^i}{R_p E_2^i},
\end{align}
where $C_{rs}^i\triangleq  \int_{T_1^i}^{T_2^i} r_i(t) s_i(t-\hat{\tau}_i) dt$.
Then, based on the acquired TOA estimates, the TDOA estimates can be calculated as (see \eqref{eq:tdoadef})
\begin{align}\label{eq:TDOAest}
\hat{d}_i = \hat{\tau}_i-\hat{\tau}_{1}\,,
\end{align}
for $i=2,\dots,N_L$. The transition from the TOA estimates to the TDOA estimates is important for reducing the computational complexity as it eliminates the need for estimating $\Delta$ since the time offset information is not present in the TDOA.

In the second step, the aim is to estimate $\boldsymbol{l}_r$ based on $\hat{d}_i$ for $i=2,\ldots,N_L$ and $\hat{\alpha}_i$ for $i=1,\ldots,N_L$. To that aim, the following proposition is presented.

\textit{\textbf{Proposition~2}:} When $E_3^i=0$ for $i=1,\dots,N_L$ and the SNR levels are sufficiently high for all optical channels (i.e., $\alpha_i^2 R_p^2E_2^i \gg \sigma^2$), $\boldsymbol{\hat{d}}\triangleq [\hat{d}_2,\ldots,\hat{d}_{N_L}]^T$ and $\boldsymbol{\hat{\alpha}}\triangleq [\hat{\alpha}_1,\ldots,\hat{\alpha}_{N_L}]^T$ can approximately be modeled as
\begin{align}\label{eq:dapprox}
&\boldsymbol{\hat{d}} =
\boldsymbol{d}+
\boldsymbol{\eta}\,,\\
\label{eq:alphaapprox}
&\boldsymbol{\hat{\alpha}} =
\boldsymbol{\alpha}+
\boldsymbol{\zeta}\,,
\end{align}
where $\boldsymbol{d}\triangleq [d_2,\dots,d_{N_L}]^T$, $\boldsymbol{\alpha}\triangleq [\alpha_1,\dots,\alpha_{N_L}]^T$, $\boldsymbol{\eta}$ is a zero mean Gaussian random vector with covariance matrix
\begin{gather}\label{eq:covMatd}
\boldsymbol{\Sigma}_{\boldsymbol{d}}= \boldsymbol{1} \frac{\sigma^2}{R_p^2\alpha_1^2 E_1^1} + \frac{\sigma^2}{R_p^2} \,\mathrm{diag}\left(\frac{1}{\alpha_2^2 E_1^2},\ldots,\frac{1}{\alpha_{N_L}^2 E_1^{N_L}}\right)
\end{gather}
with $\boldsymbol{1}$ denoting a matrix of all ones and $\mathrm{diag}(\cdot)$ representing a diagonal matrix, and $\boldsymbol{\zeta}$ is a zero mean Gaussian random vector with covariance matrix
\begin{gather}\label{eq:covMata}
\boldsymbol{\Sigma}_{\boldsymbol{\alpha}}= \frac{\sigma^2}{R_p^2}\,\mathrm{diag}\left(\frac{1}{E_2^1},\ldots,\frac{1}{E_2^{N_L}}\right).
\end{gather}
Furthermore, $\boldsymbol{\eta}$ and $\boldsymbol{\zeta}$ are independent.

\textit{Proof:} Please see Appendix \ref{sec:prooflemma1}.

It is important to note that the assumption that $E_3^i=0$ is not a significant limitation for common practical applications since $E_3^i=0.5(s_i^2(T_s)-s_i^2(0))$ and common pulses employed in practice satisfy $s_i(T_s)=s_i(0)$.

The results in Proposition~2 can be explained and utilized for localization as follows: As the estimates $\hat{d}_i$ and $\hat{\alpha}_i$ from the first step are optimal in the ML sense, those estimates should be asymptotically unbiased and efficient \cite{poor_book}. Then, the approximate models in \eqref{eq:dapprox}--\eqref{eq:covMata} can be used to estimate $\boldsymbol{l}_r$ by considering the ML parameter estimation framework. It is noted that the only parameter to be estimated now is $\boldsymbol{l}_r$, which is included both in $\boldsymbol{\alpha}$ and $\boldsymbol{d}$. Therefore, considering the approximate models in \eqref{eq:dapprox}--\eqref{eq:covMata}, the log-likelihood function of the estimates $\boldsymbol{\nu} \triangleq [\boldsymbol{\hat{d}}^T,\,\boldsymbol{\hat{\alpha}}^T]^T$ from the first step can be written as
\begin{align}\label{eq:logliketwostep}
\Gamma(\boldsymbol{\nu})
= -\frac{1}{2} \log |2\pi \boldsymbol{\Sigma}|
-\frac{1}{2}
\left(
(\boldsymbol{\nu}-\boldsymbol{\mu})^T
\boldsymbol{\Sigma}^{-1}
(\boldsymbol{\nu}-\boldsymbol{\mu})
\right),
\end{align}
where
$\boldsymbol{\Sigma}\triangleq \mathrm{Diag}(
\boldsymbol{\Sigma}_{\boldsymbol{d}},
\boldsymbol{\Sigma}_{\boldsymbol{\alpha}}
)$ with $\mathrm{Diag}(\cdot)$ denoting a block diagonal matrix of its arguments,
$\boldsymbol{\mu}\triangleq [\boldsymbol{d}^T,\,\boldsymbol{\alpha}^T]^T$,
and $\log$ denotes the natural logarithm. Based on the log-likelihood function in \eqref{eq:logliketwostep}, the ML estimate of $\boldsymbol{l}_r$ can be written as
\begin{align}\label{eq:ts_ml}
\boldsymbol{\hat{l}}_r=
\argmin_{\boldsymbol{l}_r}\,
\log|\boldsymbol{\Sigma}_{\boldsymbol{d}}|
+
(\boldsymbol{\nu}-\boldsymbol{\mu})^T
\boldsymbol{\Sigma}^{-1}
(\boldsymbol{\nu}-\boldsymbol{\mu})
\end{align}
which is the estimator in the second step of the proposed two-step estimator.

It is worth noting that the first term in \eqref{eq:ts_ml} does not contain $\boldsymbol{\Sigma}_{\boldsymbol{\alpha}}$ since the $\log|2\pi \boldsymbol{\Sigma}|$ term in \eqref{eq:logliketwostep} can be written as the summation of individual determinants and $\boldsymbol{\Sigma}_{\boldsymbol{\alpha}}$ does not depend on the unknown parameters, namely $\boldsymbol{l}_r$. On the other hand, $\boldsymbol{\Sigma}_{\boldsymbol{d}}$ depends on $\boldsymbol{l}_r$ through $\alpha_i$'s and therefore it is present in the objective function in \eqref{eq:ts_ml}. In addition, it should be emphasized that the covariance matrix $\boldsymbol{\Sigma}$ is not diagonal due to the transition from the TOA to the TDOA measurements, which results in correlations among the noise components in the TDOA estimates.

To summarize, the proposed two-step estimator works as follows: First, the ML estimates of $\alpha_i$ and $\tau_i$ are obtained from \eqref{eq:corr} and \eqref{eq:alpEst} for $i=1,\dots,N_L$. Then, from $\tau_i$'s, the TDOA parameters, $d_i$'s, are computed as in \eqref{eq:TDOAest} by selecting one of the LED transmitters as the reference. In the second step, the approximate models for the TDOA and RSS estimates obtained in Proposition~2 are utilized, which leads to the estimator in \eqref{eq:ts_ml} for the location of the VLC receiver.

It is important to compare the direct positioning approach in the previous section with the proposed two-step estimator. The direct positioning approach leads to the optimal $\boldsymbol{l}_r$ that maximizes the log-likelihood function in \eqref{eq:loglike}. On the other hand, the two-step estimator first maximizes the individual log-likelihood functions related to the received signals due to different LED transmitters, and obtains the optimal TOA and RSS estimates in the ML sense. Then, to remove the effects of the time offset, the TDOA estimates are generated from the TOA estimates. Then, given the RSS and the TDOA estimates, the second step employs the ML position estimator for high SNR scenarios. Therefore, the suboptimality of the two-step approach is related to both the TDOA generation operation and the suboptimality of the estimator in the second step when SNRs are not sufficiently high.

Regarding the computational complexity, the optimization problem in \eqref{eq:mle} corresponding to the direct positioning approach involves a search over a four-dimensional space as the optimization variable contains both $\boldsymbol{l}_r$ and $\Delta$. On the other hand, the two-step estimator does not take $\Delta$ as an argument of the objective functions neither in the first step (see \eqref{eq:corr} and \eqref{eq:alpEst}) nor in the second step (see \eqref{eq:ts_ml}). In particular, the two-step estimator requires $N_L$ one-dimensional optimizations as in \eqref{eq:corr} and one three-dimensional optimization as in \eqref{eq:ts_ml}. Hence, the use of the two-step estimator is advantageous over the direct estimator in terms of computational complexity.

%================================================================================================

\section{Numerical Results}\label{sec:nume}

This section provides numerical examples to illustrate the theoretical limits on localization and the performance of the positioning algorithms in the previous section. A room with a width and a depth of $15$ m and a height of $4$ m is considered. Four LED transmitters are placed at locations $\boldsymbol{l}_t^1=[10,10,4]^T$, $\boldsymbol{l}_t^2=[5,10,4]^T$, $\boldsymbol{l}_t^3=[10,5,4]^T$ and $\boldsymbol{l}_t^4=[5,5,4]^T$ m, and they point downwards, i.e., $\boldsymbol{n}_t^i=[0,0,-1]^T$ for $i=1,2,3,4$. The VLC receiver is located on the floor and it points upwards, i.e., $\boldsymbol{n}_r=[0,0,1]^T$. {In addition, it is assumed that there are no wall reflections in the room and only LOS components of the transmitted signals are received by the VLC receiver (see Section~\ref{sec:sysmodel}).}

A VLP system similar to the ones employed in \cite{PosAccuracyTOA,comparative_keskin} is considered in the following simulations. Namely, the responsivity and the area of the photo detector are taken as $R_p=0.4$ {mA/mW} and $A_r=1$ $\text{cm}^2$, respectively. The Lambertian order of the LED transmitter is given by $m=1$. Moreover, the power spectral density level of the AWGN is set to $\sigma^2=1.336\times 10^{-22}$ {W/Hz}. Also, the transmitted signal from the $i$th transmitter is modeled as
\begin{align}\label{eq:sioft}
s_i(t) = A(1+\cos(2\pi f_c t-\pi )), \quad t\in[0,T_s]
\end{align}
with $f_cT_s\in\mathbb{Z}$, where $f_c$ is the center frequency and $A$ denotes the source optical power that is used to set the SNR value of the corresponding optical channel. Moreover, by plugging the signal in \eqref{eq:sioft} into \eqref{eq:e1} and \eqref{eq:e2}, one can obtain that $E_1^i=\frac{4}{3}\pi^2f_c^2 E_2^i$ and $E_2^i=\frac{3}{2}A^2T_s$. {Under this signal model, the CRLB for estimating $\alpha_i$ and $\tau_i$ can be obtained by inserting these $E_1^i$ and $E_2^i$ values into \eqref{eq:crlbalphatau}, which consequently yields
\begin{align}
&\mathrm{Var}(\hat{\alpha}_i) \geq \frac{2\sigma^2}{3R_p^2 A^2T_s} \label{eq:crlbalpha}, \\
&\mathrm{Var}(\hat{\tau}_i) \geq \frac{\sigma^2}{2\pi^2 \alpha_i^2 R_p^2 f_c^2 A^2T_s}.
\label{eq:crlbtau}
\end{align}
}
Notice also that the signal in \eqref{eq:sioft} satisfy $E_3^i=0$, which is employed in the two-step positioning approach.

{In the following, we first focus on two-dimensional positioning in which $l_{r,3}$, i.e., the height of the VLC receiver, is known. Next, we also investigate three-dimensional positioning in which all the coordinates of $\boldsymbol{l}_r$, in addition to the time offset $\Delta$, are unknown parameters.}

\subsection{Two-Dimensional Positioning}

\begin{figure}
\centering
\includegraphics[width=5in]{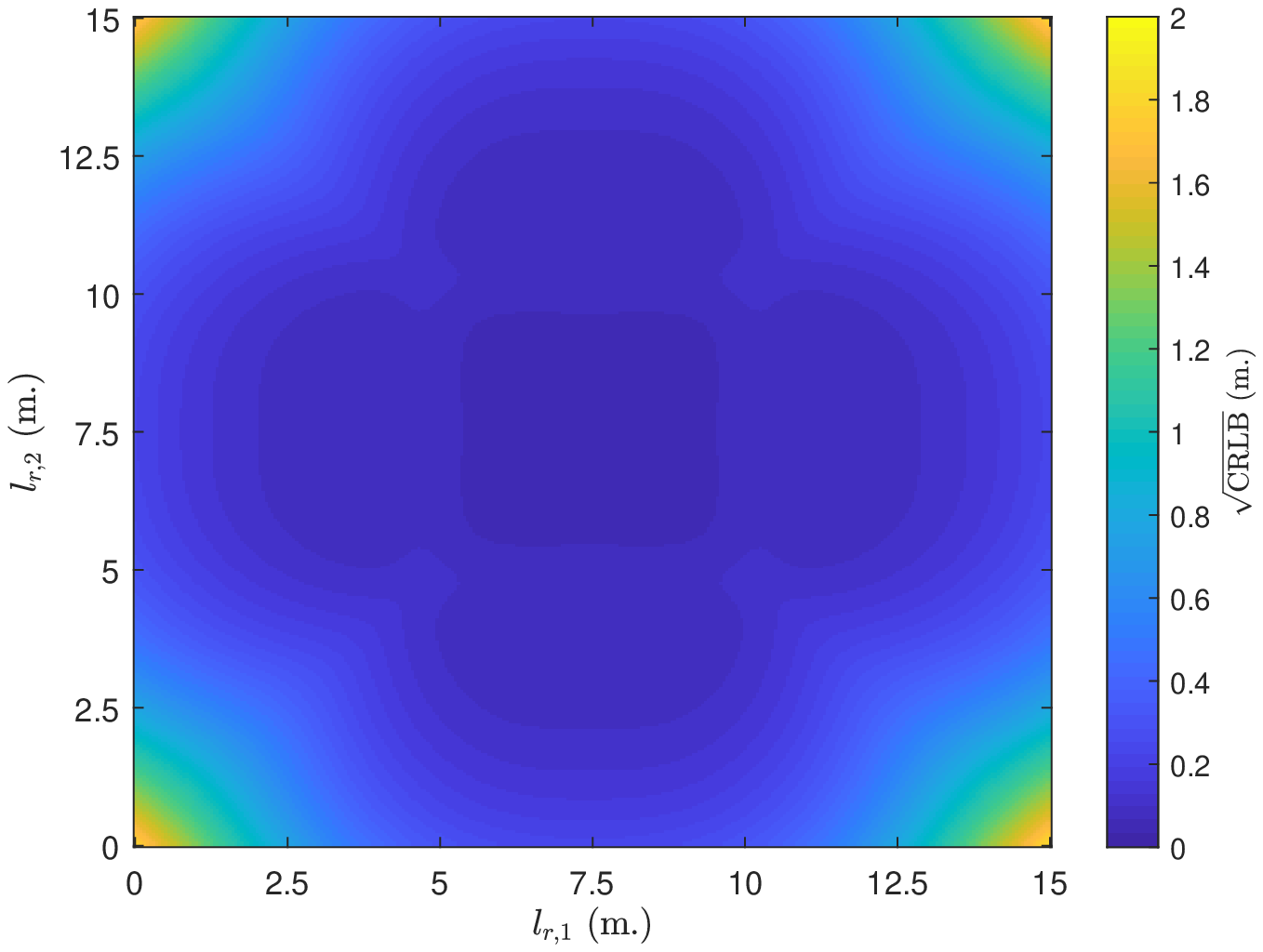}
\caption{{$\sqrt{\mathrm{CRLB}}$ when the VLC receiver moves within the room on the floor, where $f_c=100$ MHz, $T_s=10^{-6}$ s, and $A=1\,$W.}}
\label{fig1:surface}
\end{figure}

Based on the theoretical limits on positioning, the effects of various parameters are investigated for quasi-synchronous VLP systems in the following. First, the CRLBs are computed when the VLC receiver moves within the room on the floor in order to illustrate how the CRLB is affected by the location of the VLC receiver. In Fig.~\ref{fig1:surface}, {the square-root of} the CRLBs are plotted versus the first and the second coordinate of the position of the VLC receiver, where $l_{r,3}=0$, $f_c=100$ MHz, $T_s=10^{-6}$ s, and $A=1\,$W. It can be observed that the CRLB increases significantly towards the corners of the room. This is expected since the received signal powers at the VLC receiver due to the signals coming from the LED transmitters, except for the one that is closest, reduce significantly towards the corners, which can be verified by \eqref{eq:alphadef}. Therefore, the bound on the positioning accuracy increases significantly as the VLC receiver can utilize only the signal coming from the LED transmitter that is closest for determining its position. On the other hand, when $l_{r,1}$ and $l_{r,2}$ ranges in the interval $[5,10]\,$m meaning that the VLC receiver is inside the region restricted by the positions of the LED transmitters, {the square-root of} the CRLB is on the order of $0.1\,$m or lower, which is much smaller than the CRLBs at the corners. By utilizing the signals coming from more than one LED transmitter, it is possible to estimate the position of the VLC receiver more accurately in this case. All in all, in practical applications, the number of LED transmitters and their locations should be set based on the room dimensions in order to achieve high accuracy at all places in the room.

\begin{figure}
\centering
\includegraphics[width=5in]{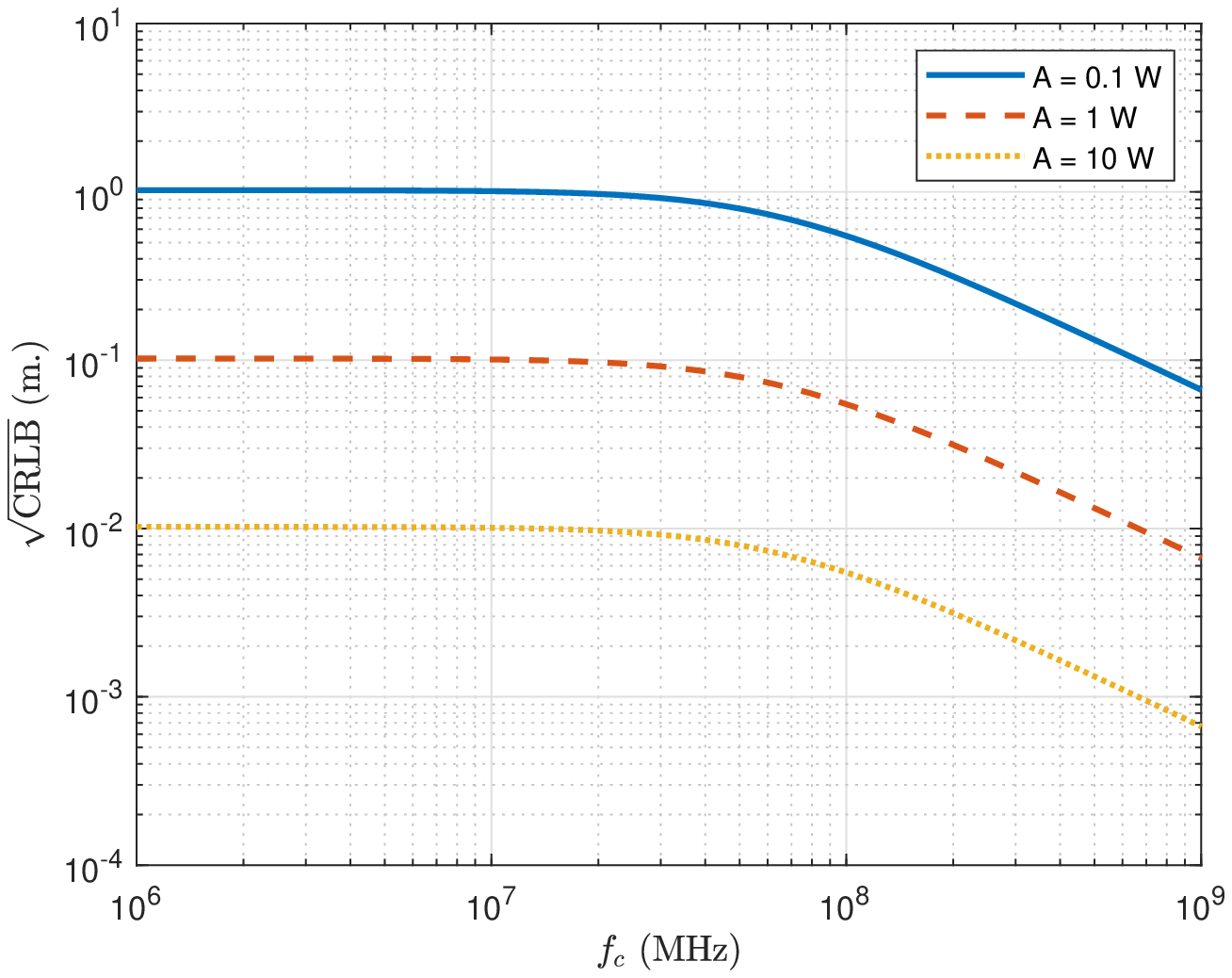}
\caption{{$\sqrt{\mathrm{CRLB}}$ versus $f_c$ for $\boldsymbol{l}_r=[6,5.75,0]^T\,$m., and $T_s=10^{-6}$ s.}}
\label{fig2:fcvscrlb}
\end{figure}

Secondly, {the square-root of} the CRLB is plotted versus $f_c$ in Fig.~\ref{fig2:fcvscrlb} for $A=0.1\,$W, $A=1\,$W, and $A=10\,$W, where $T_s=10^{-6}\,$s and $\boldsymbol{l}_r=[6,5.75,0]\,$m. It is noted that increasing the center frequency does not provide any gains in the positioning accuracy for low frequencies. On the other hand, there is a critical frequency after which an increase in the center frequency improves the positioning accuracy. The intuition behind this observation is as follows: For low frequencies, the integral term in \eqref{eq:mle}, equivalently the T(D)OA parameter, does not carry significant information and the positioning is performed mainly based on the RSS information.
% Moreover, by looking at the CRLB expression of the channel attenuation factor in \eqref{eq:crlbalphatau}, it can be seen that its variance is lower bounded by $\sigma^2/(R_p^2E_2^i)$ with $E_2^i=3A^2T_s/2$; hence, it does not depend on $f_c$.
{Moreover, it is seen that the CRLB expression of the channel attenuation factor in \eqref{eq:crlbalpha} does not depend on $f_c$.}
Since in the low frequencies the RSS information is utilized and this information does not depend on $f_c$, the CRLB remains the same with respect to $f_c$. On the other hand, for high frequencies the T(D)OA information is also utilized.
%As the variance of the TOA parameter is lower bounded by $\sigma^2/(R_p^2E_1^i\alpha_i^2)$ with $E_1^i=\frac{4}{3}\pi^2f_c^2 E_2^i$, the CRLB starts decreasing with $f_c$ for high frequencies.
{Since the lower bound on the variance of the TOA parameter is inversely proportional to $f_c^2$, as can be deduced from \eqref{eq:crlbtau}, the CRLB starts decreasing with $f_c$ for high frequencies.}

\begin{figure}
\centering
\includegraphics[width=5in]{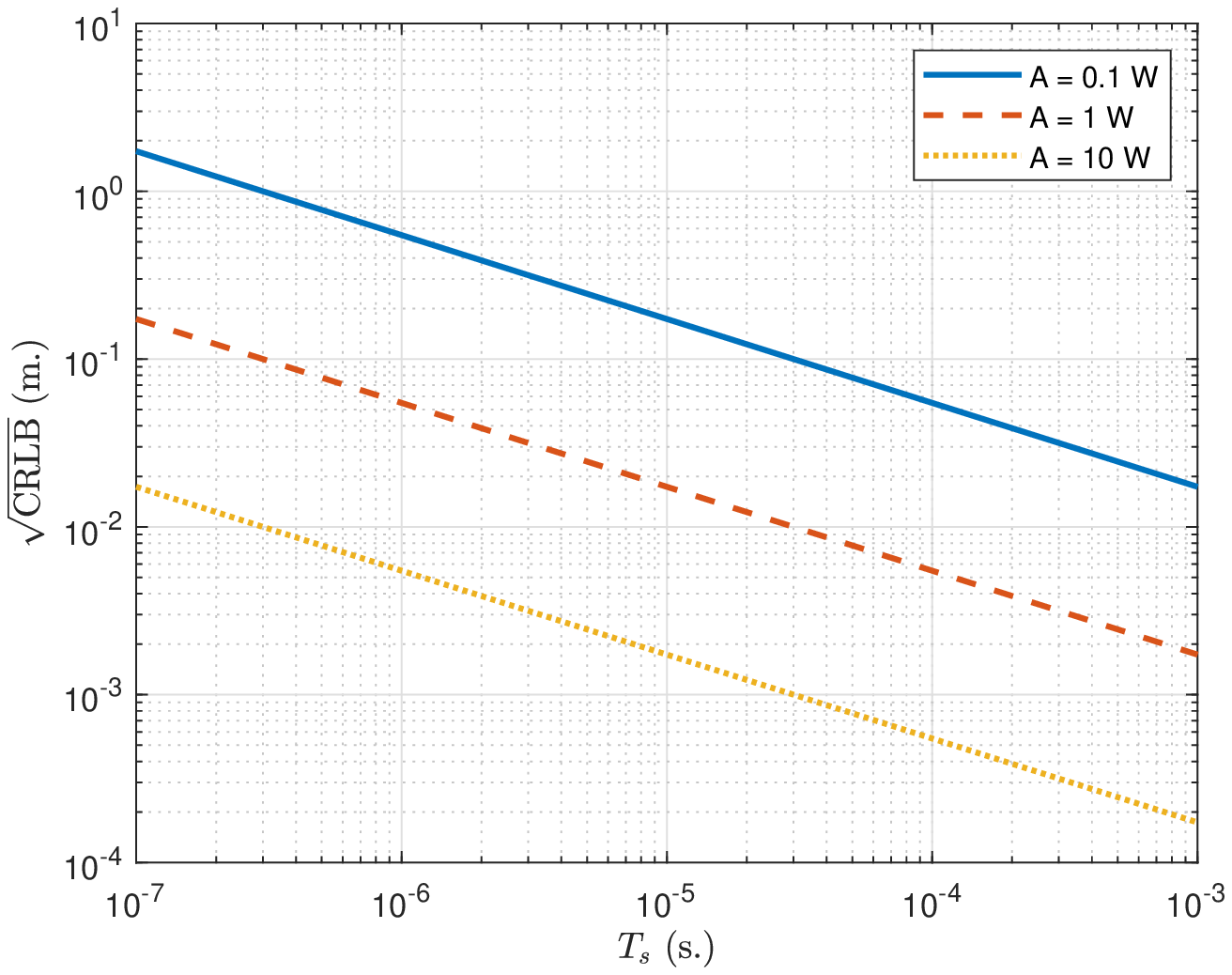}
\caption{{$\sqrt{\mathrm{CRLB}}$ versus $T_s$ for $\boldsymbol{l}_r=[6,5.75,0]^T\,$m., and $f_c=100\,$MHz.}}
\label{fig3:Tsvscrlb}
\end{figure}

In addition, the effects of $T_s$ on the CRLB are investigated in Fig.~\ref{fig3:Tsvscrlb} for $\boldsymbol{l}_r=[6,5.75,0]\,$m and $f_c=100\,$MHz. It is observed that the CRLB decreases with $T_s$ for all optical power levels, which is expected since both the RSS and T(D)OA information increases with $T_s$, as can be deduced from their corresponding CRLB expressions in {\eqref{eq:crlbalpha} and \eqref{eq:crlbtau}}. Therefore, regardless of the type of information that is used, increasing $T_s$ improves the positioning accuracy.

\begin{figure}
\centering
\includegraphics[width=5in]{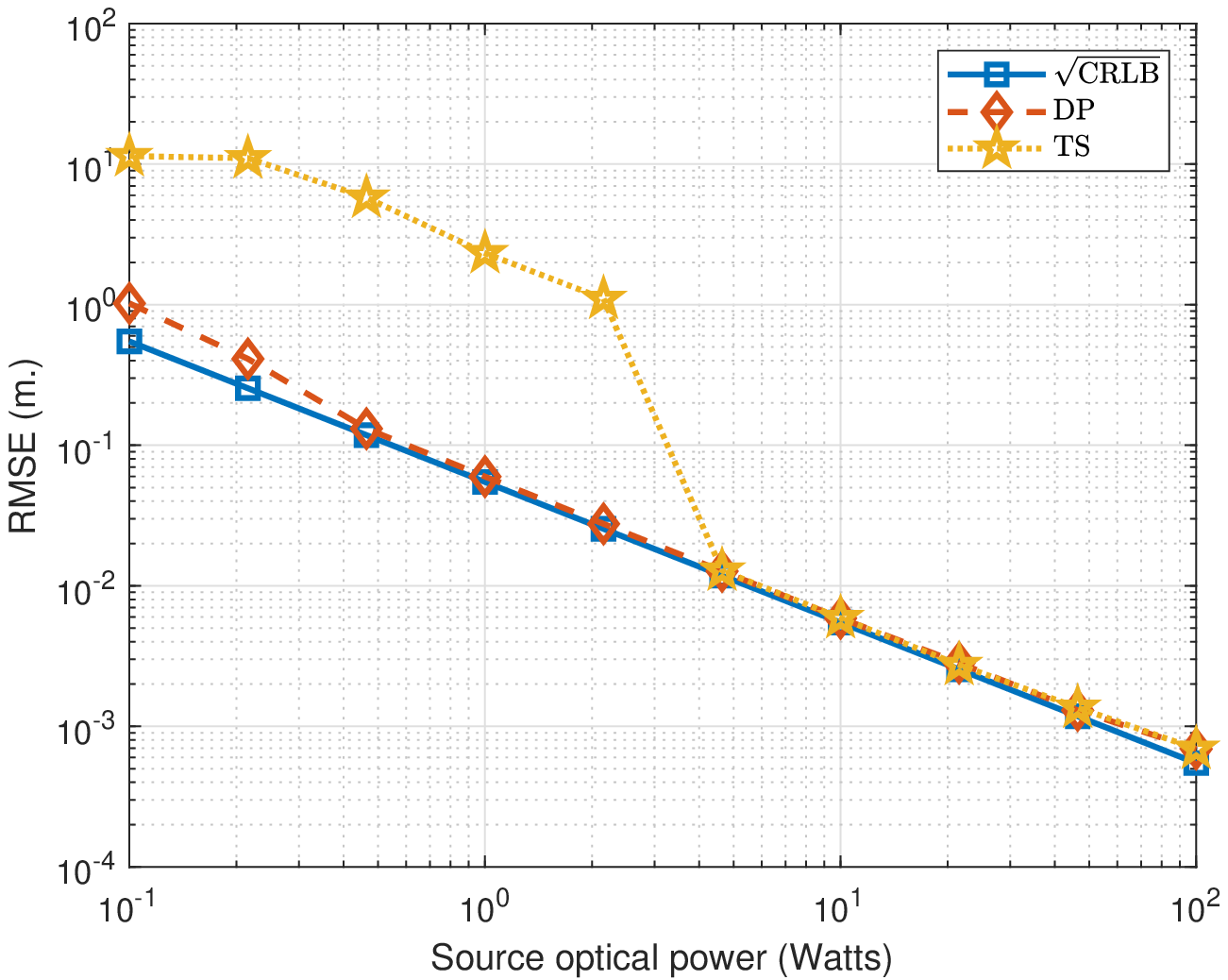}
\caption{{RMSE versus source optical power, where $f_c=100\,$MHz, $\boldsymbol{l}_r=[6,5.75,0]^T\,$m., and $T_s=10^{-6}\,$s.}}
\label{fig4:Avsrmse}
\end{figure}

\begin{figure}
\centering
\includegraphics[width=5in]{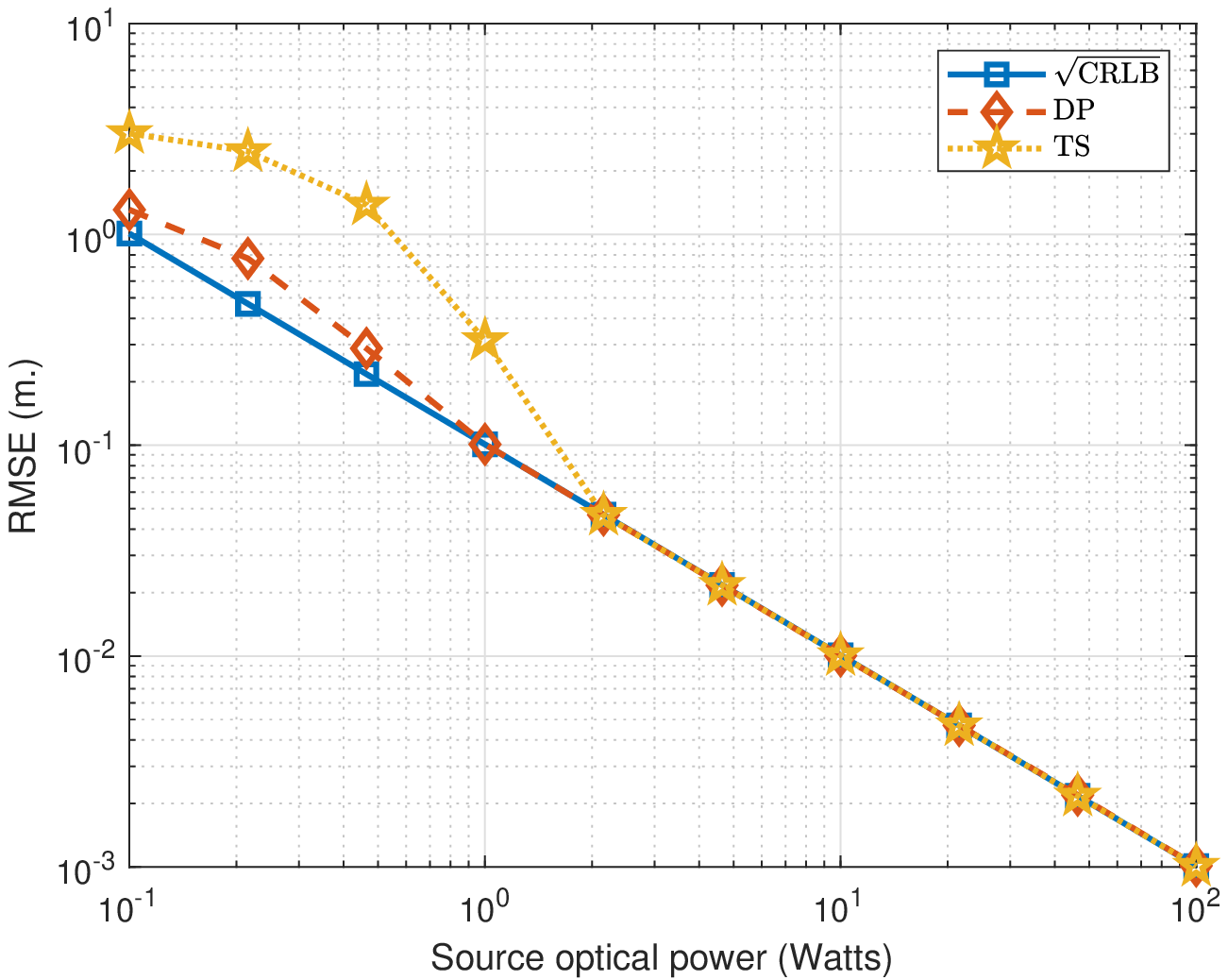}
\caption{{RMSE versus source optical power, where $f_c=10\,$MHz, $\boldsymbol{l}_r=[6,5.75,0]^T\,$m., and $T_s=10^{-6}\,$s.}}
\label{fig5:Avsrmse}
\end{figure}

Next, the ML estimators developed in Section~\ref{sec:mle}, namely, the direct positioning and two-step positioning approaches, are implemented. Their root mean-squared-error (RMSE) performance is compared against the square-root of the CRLB, which provides a lower bound on MSE of any unbiased estimator. First, the RMSE versus the source optical power is plotted in Fig.~\ref{fig4:Avsrmse}, where $\boldsymbol{l}_r=[6,5.75,0]\,$m, $f_c=100\,$MHz, and $T_s=10^{-6}\,$s. As expected, increasing the source optical power and consequently the SNR level decreases the CRLB. By looking at the entries of the FIM in \eqref{eq:Jmn}, \eqref{eq:Jk4}, and \eqref{eq:J44}, it can be verified that the FIM is proportional to $A$ in case of $E_3^i=0$ as both $E_1^i$ and $E_2^i$ are proportional to $A$. Hence, the CRLB becomes inversely proportional to $A$. Moreover, the RMSE performance of the direct positioning approach becomes comparable to the CRLB especially for mid-to-high SNR levels. Hence, the asymptotic optimality of the direct positioning approach is verified in this specific scenario. The two-step positioning technique also achieves an accuracy level as high as the CRLB for high SNR levels. Thus, it is justified for this scenario that when the high SNR assumption holds, the two-step estimator also approaches the optimal ML estimator. On the other hand, for low SNRs, there is significant degradation in the positioning accuracy. The reason behind this phenomenon is as follows: The first step of the two-step positioning approach aims to find the TOA values for which the correlator output is maximized, as stated in \eqref{eq:corr}. When the SNR level is low, the maximizing argument can be detected around the wrong peak of the correlator output due to the noise, which results in a large difference between the true value and the estimate of the TOA parameter. As a result, the T(D)OA and consequently RSS parameters cannot be estimated accurately in the first step; hence, based on erroneous estimates, the second step results in degraded localization performance. All in all, the SNR range in which the VLC positioning system operates is important in determining whether to employ the direct estimator or the two-step estimator. If the SNR value is sufficiently large, it would be preferable to use the two-step approach due to its lower computational complexity.

Moreover, the RMSE versus the source optical power is plotted in Fig.~\ref{fig5:Avsrmse} for another scenario in which $\boldsymbol{l}_r=[6,5.75,0]\,$m, $f_c=10\,$MHz, and $T_s=10^{-6}\,$s. Similar remarks to those in the $f_c=100\,$MHz case can also be made in this case. It should be added that in the cases of $f_c=10\,$MHz and $f_c=100\,$MHz, the source optical power values after which the RMSE of the two-step estimator achieves the CRLB are different. Hence, whether to use the direct positioning or the two-step positioning approach should be decided based not only on the SNR level but also on the center frequency in practical applications.

\begin{figure}
\centering
\includegraphics[width=5in]{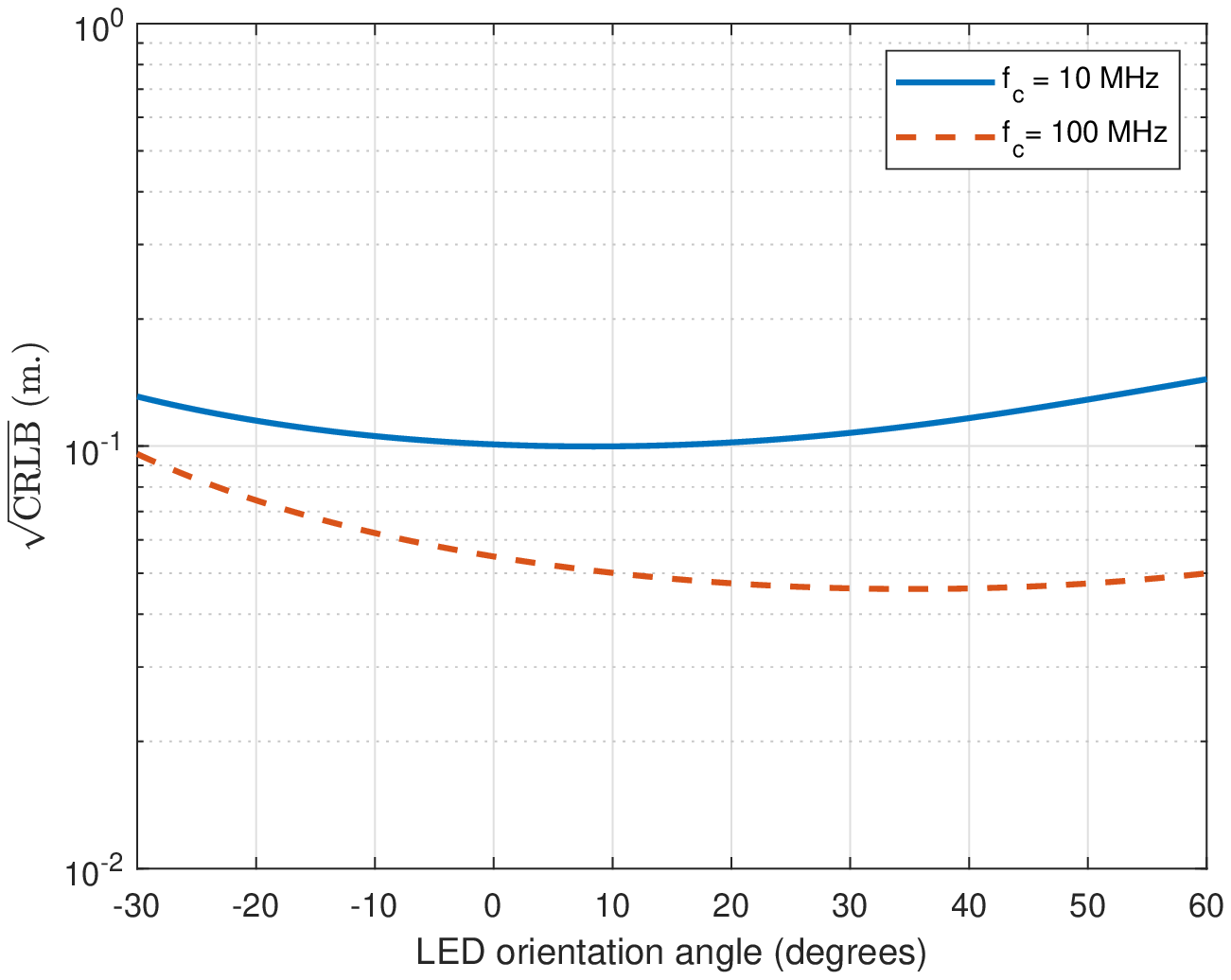}
\caption{{$\sqrt{\mathrm{CRLB}}$ versus LED orientation angle, where $T_s=10^{-6}\,$s., $\boldsymbol{l}_{r}=[6,5.75,0]^T\,$m., and $A=1\,$W.}}
\label{fig6:angle}
\end{figure}

Finally, the effects of the LED orientation on the CRLB are investigated. Note that in the previous examples, the LED transmitters look downwards while the VLC receiver looks upwards. In order to observe the effects of the orientation, the LED transmitters are tilted at an angle of $\theta$ towards the center of the room, i.e., $[7.5,7.5,4]^T\,$m. Namely, the normal vectors are given by $\boldsymbol{n}_t^1=[-n_x,-n_y,-n_z]$, $\boldsymbol{n}_t^2=[n_x,-n_y,-n_z]$, $\boldsymbol{n}_t^3=[-n_x,n_y,-n_z]$ and $\boldsymbol{n}_t^4=[n_x,n_y,-n_z]$ where $n_x=\sin(\theta)/\sqrt{2}$, $n_y=\sin(\theta)/\sqrt{2}$, and $n_z=\cos(\theta)$.
In Fig.~\ref{fig6:angle}, {the square-root of} the CRLB is plotted versus $\theta$ for $f_c=10\,$MHz and $f_c=100\,$MHz, where $T_s=10^{-6}\,$s, $A=1\,$W, and $\boldsymbol{l}_r=[6,5.75,0]\,$m. It is important to note that the best performance is not achieved in the perpendicular case, which was used in the previous simulations. However, the gain obtained by carefully adjusting the orientation is not significant. Therefore, when higher accuracy is desired for a specific scenario, it would be preferable to adjust such parameters as $T_s$, $f_c$, and $A$ rather than to fine-tune the orientation angle.

\subsection{Three-Dimensional Positioning}

\begin{figure}
\centering
\includegraphics[width=5in]{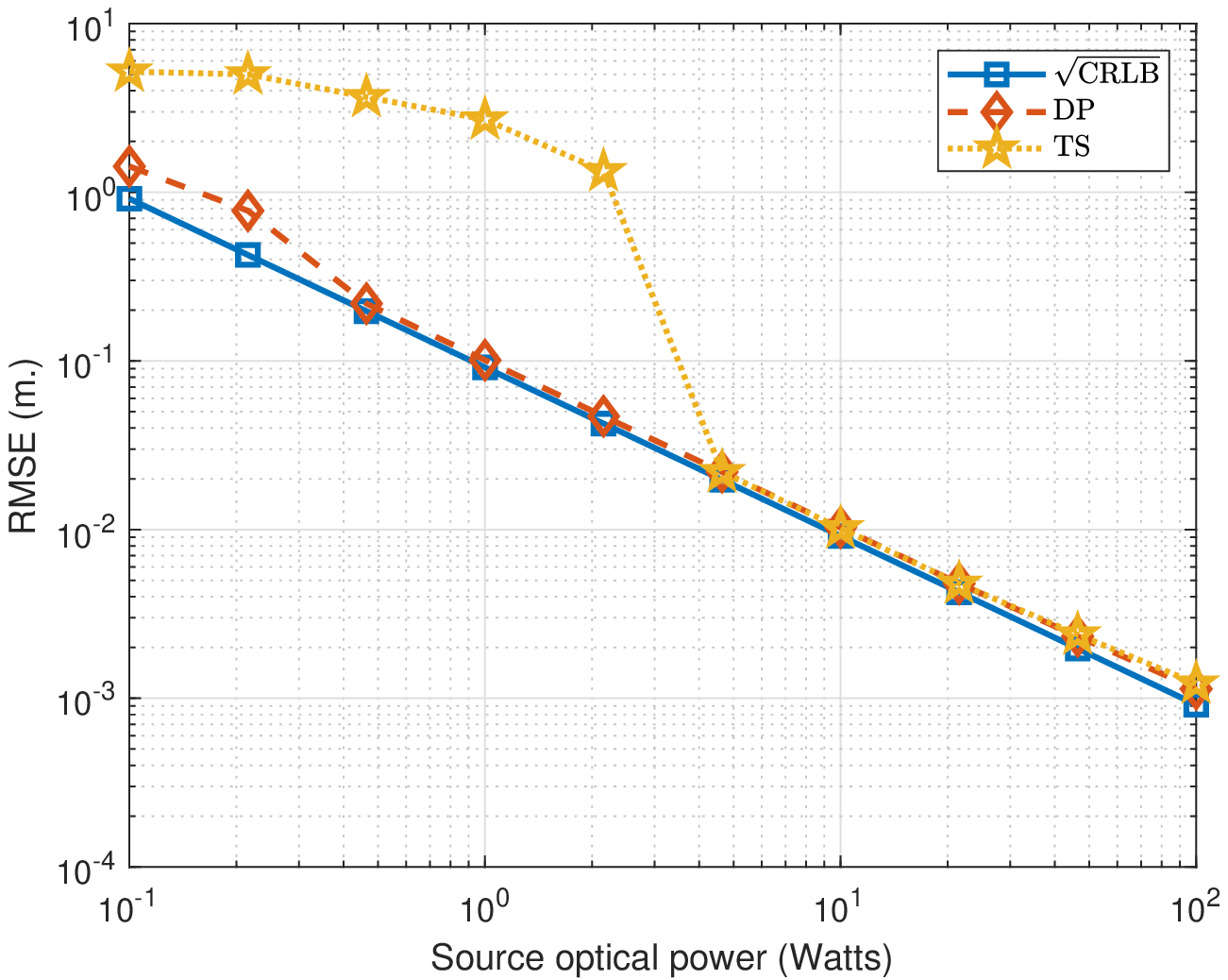}
\caption{{RMSE versus source optical power, where $f_c=100\,$MHz, $\boldsymbol{l}_{r}=[6,5.75,0]^T\,$m., and $T_s=10^{-6}\,$s.}}
\label{fig7:Avsrmse3D}
\end{figure}

\begin{figure}
\centering
\includegraphics[width=5in]{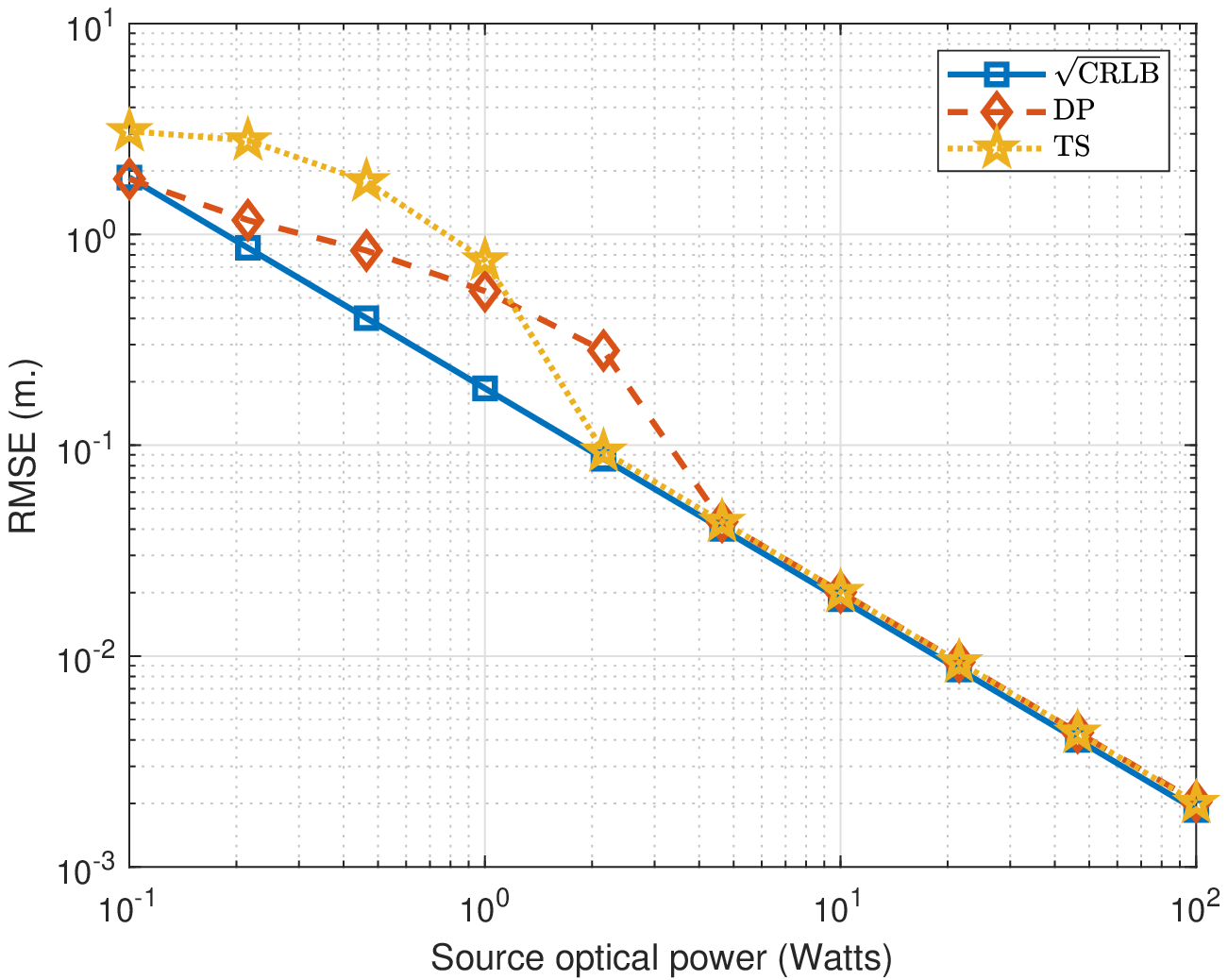}
\caption{{RMSE versus source optical power, where $f_c=10\,$MHz, $\boldsymbol{l}_{r}=[6,5.75,0]^T\,$m., and $T_s=10^{-6}\,$s.}}
\label{fig8:Avsrmse3D}
\end{figure}

{
In this part, three-dimensional positioning is considered. Namely, the height of the VLC receiver, i.e., $l_{r,3}$, is also unknown. Similar to the previous part, the VLC receiver is located at $\boldsymbol{l}_r=[6,5.75,0]\,$m., it points upwards, and the LED transmitters point downwards. Under this scenario, in Fig.~\ref{fig7:Avsrmse3D}, the RMSE performance of the proposed positioning techniques versus the source optical power is plotted together with the corresponding the theoretical limits, where $f_c=100\,$MHz and $T_s=10^{-6}\,$s. It is observed that both positioning techniques achieve accuracies that are very close to the CRLB at high SNRs. Consequently, it is observed that the theoretical limits are attained by the proposed positioning techniques at high SNRs in the three-dimensional positioning, as well. Next, we change the center frequency of the transmitted pulse to $f_c=10\,$MHz and again plot the RMSE versus the source optical power. Similar to the previous case, the performance of the positioning techniques converges to the theoretical limit in the high SNR regime. All in all, this part illustrates that both positioning techniques can achieve accuracy levels that are close to the theoretical limits for three-dimensional positioning, as well.
}

%\tcr{Bu iki grafik yeterli mi? 3D olunca Fig.~1 ve Fig.~3 icin benzer sounclar elde ediliyor. Fig.~2'de birden fazla trend gorunuyor. Bu grafikleri makale klasorunun icine koydum. }

\section{Concluding Remarks}\label{sec:conc}

In this manuscript, LED based positioning in a quasi-synchronous VLP system has been investigated. The considered system consists of LED transmitters, which emit known visible light signals, and a VLC receiver, which locates itself based on the signals coming from the LED transmitters. First, the CRLB expression has been derived for the corresponding position estimation problem. Via this expression, the effects of various system parameters on localization accuracy have been investigated. Next, ML based position estimators have been considered. In particular, the direct positioning approach, in which received signals are used directly without any intermediate steps, has been adopted. It has been observed that performance (i.e., MSE) of direct positioning converges to the theoretical limit (i.e., CRLB) at high SNRs. % for two different simulation scenarios.
Moreover, a two-step positioning technique, which is computationally efficient, has been proposed by utilizing the asymptotic properties of ML estimation. The MSE of the two-step approach closely matches with the MSE of the direct approach at high SNRs (i.e., both of them converge to the CRLB), which shows the effectiveness of the two-step approach. Furthermore, it has been observed that in the low SNR regime, the information carried in the time information (i.e., TDOA) is erroneous, distorting the overall performance of the two-step approach significantly compared to the direct approach. Hence, the two-step approach is more convenient at high SNRs due to its computational efficiency while the direct approach is more preferable in the low SNR regime due to its improved performance.

\appendix
\section{{Appendices}}

\subsection{FIM Derivation}\label{sec:fimder}

To compute the FIM, the derivatives of the likelihood function with respect to the unknown parameters are expressed first. Namely, the derivative of the log-likelihood function in \eqref{eq:loglike} with respect to $\Delta$ can be written as
\begin{align}\label{eq:crlb_der1}
\frac{\partial\Lambda}{\partial\Delta} =
-\frac{R_p}{\sigma^2}
\sum_{i=1}^{N_L}
\int_{T_{1,i}}^{T_{2,i}}
n_i(t)
\alpha_i s_i'(t-\tau_i)
dt
\end{align}
and with respect to $l_{r,k}$ as
\begin{align}\label{eq:crlb_der2}
\frac{\partial\Lambda}{\partial l_{r,k}} =
\frac{R_p}{\sigma^2}
\sum_{i=1}^{N_L}
\int_{T_{1,i}}^{T_{2,i}}
n_i(t)
\bigg(
\frac{\partial\alpha_i}{\partial l_{r,k}}
s_i(t-\tau_i) -
\alpha_i s_i'(t-\tau_i)
\frac{\partial\tau_i}{\partial l_{r,k}}
\bigg)
dt\,.
\end{align}
Then, plugging \eqref{eq:crlb_der1} and \eqref{eq:crlb_der2} into \eqref{eq:fim} yields the results stated in \eqref{eq:Jmn}, \eqref{eq:Jk4}, and \eqref{eq:J44}.

\subsection{Proof of Proposition~2}\label{sec:prooflemma1}

{
Consider the estimation of $\tau_i$ and $\alpha_i$ based on the received signal from the $i$th LED transmitter, $r_i(t)$. In \cite[App.~A]{Keskin_Direct}, it is shown that when $E_3^i=0$, the inverse of the FIM for estimating $\alpha_i$ and $\tau_i$ based on $r_i(t)$ can be calculated as
\begin{align}\label{eq:crlbalphatau}
\mathbf{J}_i^{-1} =
\frac{\sigma^2}{R_p^2}
\begin{bmatrix}
1/E_2^i & 0 \\
0 & 1/(\alpha_i^2E_1^i)
\end{bmatrix},
\end{align}
where $E_1^i$ and $E_2^i$ are as in \eqref{eq:e1} and \eqref{eq:e2}, respectively.}

{By exploiting the asymptotic unbiasedness and efficiency properties of ML estimation, it can be inferred that at high SNRs, $\hat{\alpha}_i$ is a Gaussian random variable with mean $\alpha_i$ and variance ${\sigma^2}/({R_p^2 E_2^i})$, and $\hat{\tau}_i$ is a Gaussian random variable with mean $\tau_i$ and variance ${\sigma^2}/({R_p^2 E_1^i\alpha_i^2})$ \cite{Keskin_Direct}.} In other words, $\hat{\alpha}_i$ and $\hat{\tau}_i$ can be expressed at high SNRs as $\hat{\alpha}_i=\alpha_i+\zeta_i$ and $\hat{\tau}_i=\tau_i+\kappa_i$, where $\zeta_i$ and $\kappa_i$ are independent zero-mean Gaussian random variables with variances of ${\sigma^2}/({R_p^2 E_2^i})$ and ${\sigma^2}/({R_p^2 E_1^i\alpha_i^2})$, respectively. (The independence follows due to the facts that the ML estimate is Gaussian at high SNRs and the $\mathbf{J}_i^{-1}$ in \eqref{eq:crlbalphatau} is a diagonal matrix.) In order to see that $\{\zeta_i,\kappa_i\}$ and $\{\zeta_j,\kappa_j\}$ are also independent for $i\neq j$, one can write the corresponding FIM based on $r_i(t)$ and $r_j(t)$ (by taking $\{ \alpha_i,\tau_i,\alpha_j,\tau_j\}$ as the set of unknown parameters), and employ the fact that $n_i(t)$ and $n_j(t)$ are independent. In this way, it can be shown that $\{\zeta_i\}_{i=1}^{N_L}$ and $\{\kappa_i\}_{i=1}^{N_L}$ are independent sequences, which are also independent from each other.

When the TDOA estimates are generated as in \eqref{eq:TDOAest}, the $i$th TDOA estimate can be expressed as
\begin{equation}\nonumber
\hat{d}_i=\tau_i-\tau_1+\kappa_i-\kappa_1\triangleq\tau_i-\tau_1+\eta_i\,.
\end{equation}
From the arguments in the previous paragraph, it can be shown that $\hat{d}_i \sim \mathcal{N}(\tau_i-\tau_1, \frac{\sigma^2}{R_p^2\alpha_1^2 E_1^1} +  \frac{\sigma^2}{R_p^2\alpha_i^2 E_1^i})$ for $i=2,\ldots,N_L$ at high SNRs. In addition, the covariance between $\eta_i$ and $\eta_j$ can be calculated as ${\sigma^2}/({R_p^2\alpha_1^2 E_1^1})$.

Based on all these results, $\hat{\alpha}_i$'s and $\hat{d}_i$'s can be modeled as in \eqref{eq:dapprox}--\eqref{eq:covMata}. Moreover, since $\{\kappa_i\}_{i=1}^{N_L}$ and $\{\zeta_i\}_{i=1}^{N_L}$ are independent, $\{\eta_i\}_{i=2}^{N_L}$ and $\{\zeta_i\}_{i=1}^{N_L}$ also become independent, as claimed in the proposition.
\QEDA

\bibliographystyle{IEEEtran}

\bibliography{tdoa_rss}

\end{document}